\documentclass{aa}
\usepackage{graphicx}
\begin{document}
   \title{Box- and peanut-shaped bulges}

   \subtitle{III. A new class of bulges: Thick Boxy Bulges.}

   \author{R. L\"utticke
          \inst{1}\fnmsep\inst{2},
          M. Pohlen
          \inst{1}\fnmsep\inst{3},
          \and
          R.-J. Dettmar
          \inst{1}
          }

   \offprints{R. L\"utticke}

   \institute{Astronomical Institute, Ruhr-Universit\"at Bochum,
              D-44780 Bochum, Germany
         \and
             Department of Computer Science, Intelligent Information and Communication Systems, FernUniversit\"at Hagen, 
             Universit\"atsstr. 1, D-58084 Hagen, Germany\\
             \email{rainer.luetticke@fernuni-hagen.de}
          \and Instituto de Astrofisica de Canarias, La Laguna, 
           E-38200 Tenerife, Spain
             }

   \date{Received 29 November 2001 / Accepted 19 December 2003}

   \abstract{
Inspecting all 1224 edge-on disk galaxies larger than 2$\arcmin$ 
in the RC3
on Digitized Sky Survey (DSS) images (L\"utticke et al. \cite{lue2000a})
we have found several galaxies with extraordinary bulges meeting
two criteria:
They are box shaped and large in respect to the diameters of their galaxies.
These bulges are often disturbed, show
frequently prominent irregularities and asymmetries, and some
possess possible merger remnants or merging satellites.
For these bulges we have introduced the term ``Thick Boxy Bulges'' (TBBs).
About 2\,\% of all disk galaxies (S0-Sd),  
respectively 4\,\% of all galaxies with box- and peanut-shaped (b/p) bulges,
belong to this class of galaxies.
Using  multicolour CCD and NIR data we have enlarged and followed up
our sample of nearly 20 galaxies with a TBB.
The disturbed morphology of a large fraction of these galaxies shows
that many of the TBB galaxies are not dynamically settled.
For the TBBs the extent of the box shape seems to be too large to result
from a normal bar potential.
Therefore we conclude
that two classes of b/p bulges exist
with different origins.
While most ($\sim$\,96\,\%) b/p bulges can be
explained by bars alone (L\"utticke et al. \cite{lue2000b}),
the extended boxy structures of TBBs result
most likely from accreted material by infalling satellite
companions (soft merging). 
   }

   \maketitle
%

\section{Introduction}
Recent statistics of b/p bulges have revealed that nearly half of all
disk galaxies (S0-Sd) have bulges which differ from an elliptical shape
(L\"utticke et al. \cite{lue2000a}, hereafter Paper I).
Several studies have shown that there is strong evidence
that bar instabilities and resonances are at work in most
cases producing these b/p bulges (Combes et al. \cite{com90}, Raha et al. \cite{rah},
Kuijken \& Merrifield \cite{kui}, Bureau \& Freeman \cite{bf},
L\"utticke et al. \cite{lue2000b}, hereafter Paper II, Athanassoula \& Misiriotis \cite{am02}, Patsis et al. \cite{psa02}).
In addition to spontaneous bar instabilities in disks, 
such instabilities can also be triggered by interactions followed by accretions
of material
(Noguchi \cite{nog}, Gerin et al. \cite{ger}, Walker et al. \cite{wmh}).
The evolutionary scenario --- interaction, possible accretion, and then
the formation of a bar as progenitor of  b/p structures --- is supported
by observations and simulations (Fisher et al. \cite{fis} and Mihos et al. \cite{mih}). 
Observational evidence for accreted material in b/p bulges is also
found by Bureau \& Freeman (\cite{bf}) for a few b/p bulges in a kinematical
study.

On the other side several theories exist for the origin of b/p bulges,
explaining their non elliptical shape directly by merger events.
Binney  \& Petrou (\cite{bin}) and Rowley (\cite{row})
proposed mergers of two disk galaxies
as a formation process of b/p bulges. However, the
precise alignment of the spin and orbital angular momenta of
the two galaxies require very special conditions. Therefore such mergers would be
quite rare (Bureau \cite{bur}).
The accretion of satellite galaxies could be
a more common process to form b/p bulges
(Binney  \& Petrou \cite{bin}, Whitmore \& Bell \cite{whi}).
For such a scenario the satellite must have an oblique impact angle
(Whitmore \& Bell \cite{whi}) and
a mass, which is large enough to produce morphological changes and small
enough not to disrupt the stellar disk
(Barnes \cite{bar}, Hernquist \cite{her}).
The theories for a possible origin of b/p bulges based on accretion
events are supported by N-body simulations of
Hernquist \& Quinn (\cite{hq}).
Additionally, Dettmar (\cite{det89}) and Dettmar \& Barteldrees (\cite{db90a})
pointed out that
some boxy bulges with large radial extent
cannot be explained within an evolutionary scenario of b/p bulges
based on bars and that asymmetries and irregularities in a few galaxies
with boxy bulges rather hint to a recent merger event.

A third theory for the formation of b/p bulges has not found
further support in the literature, because the proposed
external cylindrically symmetric torques (May et al. \cite{may})
as origin of such bulges
are difficult to relate to some astrophysical
counterparts (Combes et al. \cite{com90}).

Whereas our previous study (Paper II)
revealed that the aspect angle of
bars is the main cause for the shape of bulges, 
we present here a sample of extraordinary b/p bulges
to investigate 
for the first time on a large basis of observational data
the relevance of theories of merger/accretion scenarios inducing a special class of b/p bulges.
In this way constraints on theories about secular evolution of bulges
(e.g. Combes \cite{com99}, Pfenniger \cite{pfe99})
can be found, since it is still open what roles play bar formation/destruction
and accretion events/soft mergers for secular dynamical evolution.
Detailed studies of the structure of b/p bulges are needed to
discriminate between the importance of 
the proposed building processes of b/p bulges.


\section{Data}
\subsection{Full data base}
We have compiled a list of b/p bulges in edge-on galaxies
consisting
of two parts:
\begin{enumerate}
\item
{\bf RC3-survey:} All galaxies out of the RC3
(\emph {Third Reference Catalogue of Bright Galaxies},
de Vaucouleurs et al. \cite{rc3}) with $D_{25} > 2\arcmin$
and $\log R_{25} \ge 0.35$, 0.30 for S0 galaxies respectively.\\
The RC3-survey
is complete over the whole sky and consists of 1224 galaxies ranging
from S0 to Sd ($-3.5\!<\!T\!<\!7.5$). 
These galaxies are visually inspected using the
DSS1\footnote{http://archive.eso.org/dss/dss}.
734 of them have bulges for which the shape can be classified (Paper I).
Additionally, for 76 galaxies of the RC3-survey
we obtained optical and for 56
galaxies NIR follow-up images.
\item {\bf Serendipitous survey:}
51 galaxies outside the selection criteria of the RC3-survey 
with imaging data taken from several different projects.\\
In detail this are: Two edge-on
galaxies (optical images) with an axis ratio outside the selection limits,
48 galaxies
(44 optical and four NIR images) with
a diameter smaller than $2\arcmin$, and one galaxy smaller than $2\arcmin$
and an axis ratio outside the limits (VLT image\footnote{http://www.eso.org/outreach/info-events/ut1fl/astroim-galaxy.html}).
\end{enumerate}
The observations of the optical and NIR data were obtained in several runs. 
Data reduction is presented elsewhere (optical:
Barteldrees \& Dettmar \cite{bd},
Pohlen et al. \cite{poh}, Paper I, Pohlen \cite{pohdiss}; NIR:
Paper II).

\subsection{Sample selection of the TBBs}

From this survey we have extracted a list of objects
that fall into a new category of disk galaxies hosting large and boxy
bulges.
Their morphological appearance suggests to call them ``Thick Boxy Bulges'' (TBBs).
Following the shape definitions of bulge types (Paper I) TBBs are
classified as type 2
(box-shaped bulge) or 3 (bulge is close to box-shaped, not elliptical).
Since ``normal'' bulges have a relative size of BUL/$D_{25} = 0.22$ 
(BUL = bulge length\footnote{The bulge length is marked
by an increasing light distribution compared to the exponential disk 
in a radial surface brightness profile (excluding
a possible bar).}, values derived in Paper II ranging from 0.10 to 0.42 including S0-Scd galaxies), 
we define a bulge as ``thick'' if BUL/$D_{25} > 0.5$ or
if bulge and disk component cannot be clearly separated due to the large extent
of the bulge.
Bulges of 13 galaxies out of the RC3-survey and 6 
galaxies from the serendipitous survey fall into the class of TBBs 
(Tab.~\ref{cat}).
Compared to the preliminary versions of this catalogue (Dettmar \& L\"utticke \cite{dl} and
L\"utticke \& Dettmar \cite{ld99})
some minor changes of the list of TBBs are made due to the homogeneous
definition of thickness and a reclassification of the bulge types with
data of higher quality.

\begin{table*}
\caption{Catalogue of galaxies with a TBB}
\label{cat}
\begin{center}
\begin{tabular}{l|cclrcccclcc}
\multicolumn{1}{c|}{(1)} & (2) & (3) & \multicolumn{1}{c}{(4)} & (5) & (6) & (7) & (8) & (9) & (10)& (11) & (12)\\
\multicolumn{1}{c|}{Galaxy} & RA & DEC & \multicolumn{1}{c}{Morph.}  &  $D_{25}$ & Survey  & Image & Bulge &
Asym- & $n$ & $w$ & Group \\
 &  (2000) & (2000) & Type & [$\arcsec$] & & Quality & \multicolumn{1}{c}{Type} &  metry & & & \\
\hline
ESO 013-012 &  01 07 & $-$80 18 & S0a & 165 & RCS & -- & 2 & &  & 1 & --\\
NGC  1030 &  02 40 & +18 02 & S0* & 95 & SDS &  fair & 3 & $\times$ & 2$^d$ & 2 &PPS2 173\\
NGC  1055 &  02 42 & +00 27 & Sb & 455 & RCS & fair & 2 &$\times$ & 2$^d$ & 2 & LGG 73\\
NGC  1589 &  04 31 & +00 52 & Sab & 190 & RCS & good & 2& $\times$ & 2 & 2 &LGG 117\\
UGC  3458 &  06 26 & +64 44 & Sb & 144 & RCS & good & 3 & & 3 & 2& ?\\
ESO 494-022 $^a$  & 08 06 & $-$24 49 & Sa & 134 &  RCS & -- &3 & &  & 1 & ? \\
NGC  3573 &  11 11 & $-$36 52 & S0a & 218 & RCS & good & 3 & $\times$ &  & 2 & LGG 229\\
NGC  4224 &  12 16 & +07 27 & Sa & 154 & RCS & -- & 3 & &  & 1 & LGG 281\\
ESO 506-003 &  12 22 & $-$25 05 & Sab & 85  & SDS & good &3 & & 2 & 2 & ?\\
ESO 322-100 &   12 49 & $-$41 27 & S0 & 56 & SDS & good &2 & & 2 & 1 & LGG 305\\
ESO 383-005 & 13 29 & $-$34 16 & Sbc & 213 & RCS & good & 2& $\times$ & 3 & 2& LGG 353\\
ESO 510-013 &  13 55 & $-$26 47 & Sa & 117 & SDS & good & 3& $\times$ &  & 2& --\\
NGC  5719 &  14 41 & $-$00 19 & Sab & 194 & RCS & -- &3 & $\times$ &  & 2 & LGG 386\\
UGC  9759   & 15 11 & +55 21 & S0a*$^b$ & 120 & RCS & fair & 2 &  & 2 & 1 & LGG 395\\
ESO 514-005 &  15 19 & $-$23 49 & Sa &125 & RCS & good & 2 &  & 1 & 1 & ?\\
UGC 10205 &  16 07 & +30 06 & Sa & 87& SDS & fair  & 3 & $\times$ & 4$^c$ & 3 & ?\\
IC   4745 &  18 42 & $-$64 56 & Sab & 128 & RCS & good &2 & $\times$ & 2 & 3 & LGG 422\\
IC   4757 &  18 44 & $-$57 10 & S0a & 83 & SDS & good & 2 & $\times$ & 3 & 1 & ?\\
NGC  7183 &  22 02 & $-$18 55 & S0 & 228 & RCS & -- & 3 & $\times$ & 1$^d$ & 2 & -- \\
\end{tabular}
\end{center}

Notes to the table:

Col.~(2) and (3): RA and DEC are rounded. \\
Col.~(4): Morphological types marked with an asterisk *
from Skiff (\cite{ski}), others from RC3. \\
Col.~(5): $D_{25}$ in arcsec from RC3.\\
Col.~(6): RCS: Detected in the RC3-survey.
SDS: Detected in the serendipitous survey (Sect.~2.1).\\
Col.~(7): Verified by optical CCD or NIR follow-up with
`good' or `fair' quality of the data.\\
Col.~(8): Bulge type as defined in Paper I.\\
Col.~(9): Asymmetric TBBs are marked ($\times$).\\
Col.~(10): Shape parameter of the luminosity profile measured
along the minor axis (Sect.~3.3).\\
Col.~(11): Interaction index as defined in van den Bergh et al. (\cite{vB}) (Sect.~3.4).\\
Col.~(12): Membership in a group from NED: LGG numbers from
Garcia (General study of group membership \cite{gar}),
?\,=\,galaxies are fainter than the limit of
completeness ($B_0 < 14$ mag) in the sample of Garcia (\cite{gar}) (Sect.~3.4).\\
\\
$^a$: Galaxy is affected by foreground stars due to position behind the Milky Way.\\
$^b$: S0 possible \\
$^c$: Two components can be detected. \\
$^d$: Values derived by Vergani et al. (\cite{vpld}).
\end{table*}


\section{Results}
\subsection{Morphology}
\begin{figure}[htb]
\centering
\includegraphics[angle=-90,width=9cm]{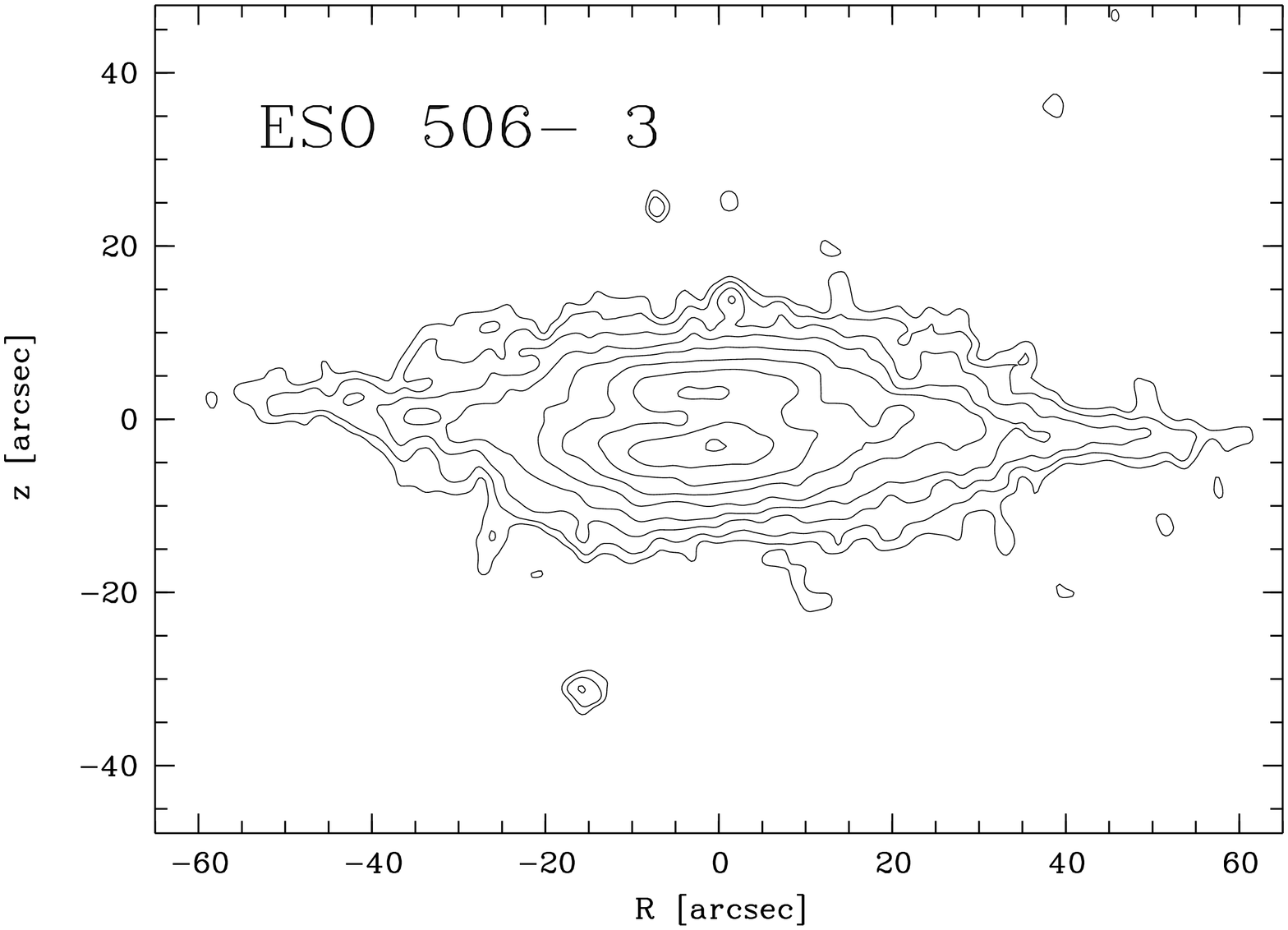}
\includegraphics[angle=-90,width=9cm]{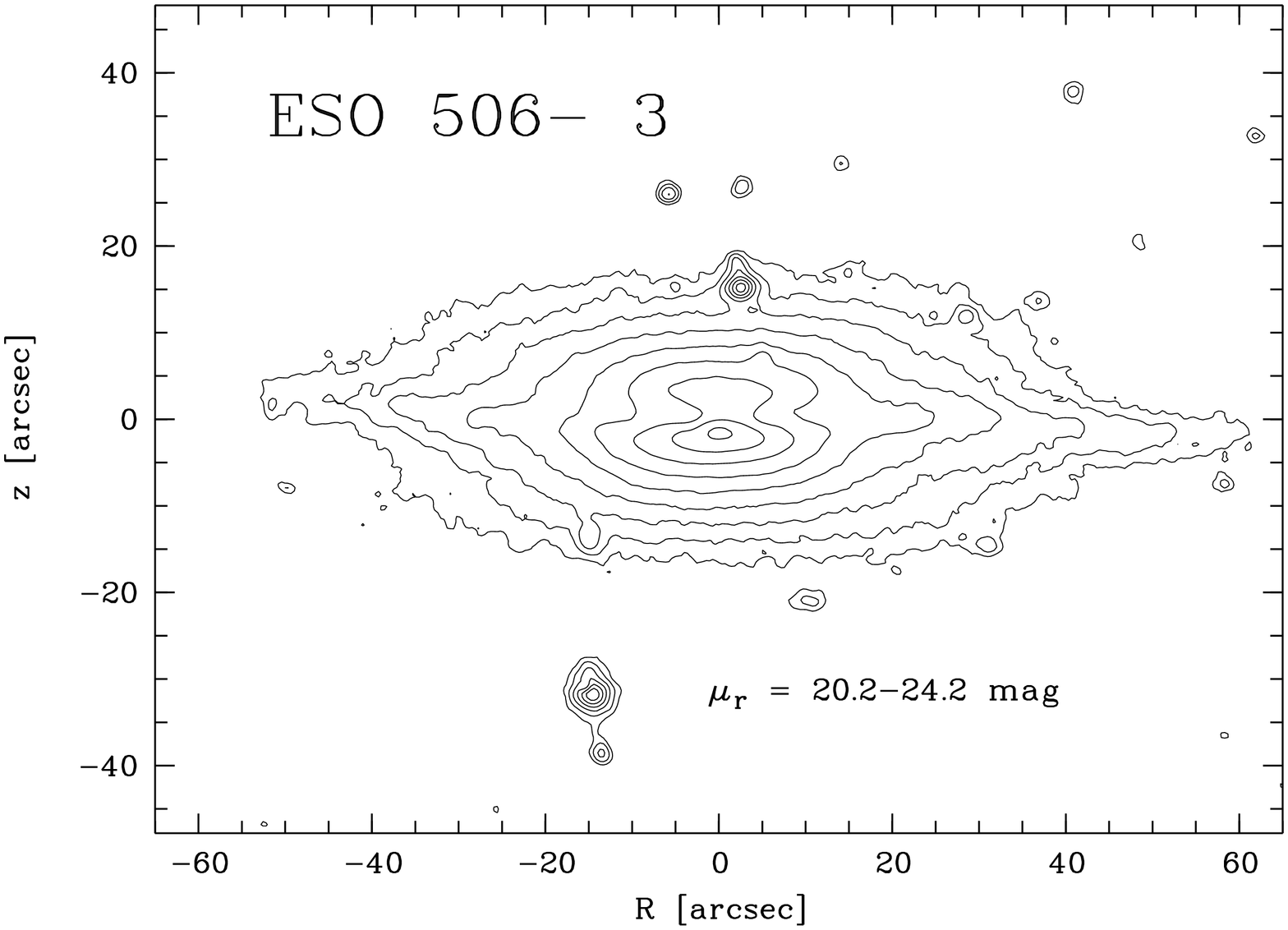}
\caption{ESO\,506-003. Top: Contour plot from DSS. Bottom: For comparison
CCD follow-up.
ESO/2.2m, 30min in $r$.}
\label{50603}
\end{figure}

TBBs show frequently extra features (e.g. twists of the
isophotes) and large scale asymmetries.
They can be divided in asymmetric (58\,\%) and rather symmetric (42\,\%)
members of the TBB class by their morphological appearance and degree of
peculiarities (Tab.~\ref{cat}).
Examples for galaxies excluded from the list of TBBs (see below: non-TBBs)
clarify the
characteristics of the TBBs.
For all DSS candidates reobserved 
(e.g. Fig.~\ref{50603}) 
the suspected peculiarities are confirmed.
It is noticeable that all galaxies with a TBB (excluding the S0 galaxy
ESO\,322-100) possess
a prominent dust lane which is often warped.

\paragraph{Asymmetric TBBs:}
The eleven asymmetric members of the class of TBBs possess a variety of individual 
features.
However, the description of their morphology in detail
depends on the quality of the data. 
Therefore it is natural 
that faint structures in galaxies investigated only with
DSS or low quality CCD data remain undetected.
The detailed description of all the peculiarities of the individual TBBs presents
a full overview about their morphology (Sect.~3.2).

\paragraph{Rather symmetric TBBs:}
There are eight out of the 19 TBBs catalogued
in Table\,\ref{cat},
which show only small asymmetries and nearly no peculiarities.
(e.g.~Fig.~\ref{50603} and Fig.~\ref{322100}).
However, they fulfill the two criteria thickness and boxiness for a TBB.
In these galaxies there is no boxy envelope visible and
the extent of the boxiness is often limited to the inner parts of the bulge.
Only the weakly asymmetric tilted disks in UGC\,3458 and ESO\,506-003 (Fig.~\ref{50603}) and
the slightly asymmetry of the merged disk/bulge component of ESO\,514-005
are remarkable.

\paragraph{Examples for non-TBBs:}
There exists three groups of galaxies which could be associated with the
class of TBB galaxies, but do not belong to this class, since they do not
fulfill the criteria of the TBBs.\\
The first group includes prominent boxy and a few peanut bulges in edge-on galaxies.
They possess
partly also irregularities, but their bulges are not large enough
to influence the shape of the whole galaxy so that disk and bulge are
separated and BUL/$D_{25}$ is smaller, although sometimes not much, than 0.5.
Such cases are e.g. NGC\,482 and IC\,4767. 
However, since decomposition of disk and bulge for S0 galaxies
is not always clear cut, BUL cannot always definitely determined.

\begin{figure}[htbp]
\centering
  \includegraphics[angle=-90,width=9cm]{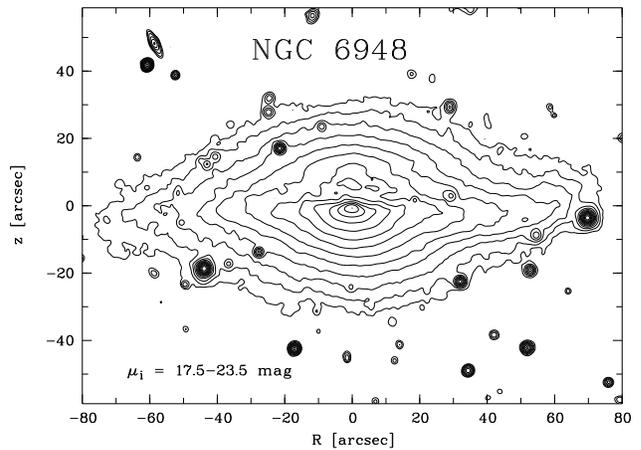}
\caption{A galaxy with a TSB:
NGC\,6948 has a prominent large bulge (BUL/$D_{25} \sim 0.6$),
but with elliptical shape.
The disk is warped, but in comparison to the asymmetric TBBs the asymmetry
is rather weak.
ESO/2.2m, 30min in $i$.}
\label{tsb}
\end{figure}

Secondly, there are thick bulges having nearly the same extent along the major
axis as the disks of these galaxies (also in
galaxies as late as Sb) and BUL/$D_{25}$ is larger
than 0.5, but these bulges are more elliptical (bulge type 4) than boxy. 
Therefore we call them TSBs (Thick Spheroidal Bulges). 
In our survey we have found five TSBs:
NGC\,4594 (Sombrero Galaxy), NGC\,6948 (Fig.~\ref{tsb}),
NGC\,7123, NGC\,7814, and ESO\,142-19. 
All these galaxies are rather symmetric.
Regarding the 734 galaxies with a classifiable bulge out of our RC3-survey
(Sect.~2.1) we get the result that 1.8\,\% of the galaxies
host a TBB and only 0.7\,\% a TSB.

\begin{figure}[htbp]
\centering
  \includegraphics[angle=-90,width=9cm]{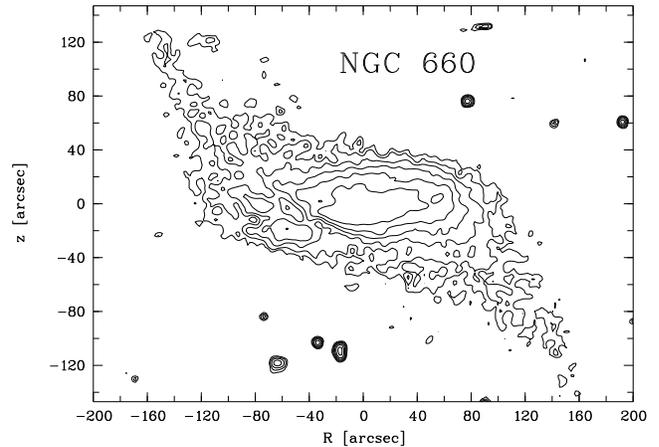}
\caption{Excluded galaxy (non-TBB):
NGC\,660 possesses boxy structures, but it is nearly disrupted. Contour
plot from DSS.}
\label{nontbb}
\end{figure}

Nearly disrupted galaxies, although having large scale boxy structures, are also
excluded from the class of galaxies with a TBB, e.g.
NGC\,660 (Fig.~\ref{nontbb}) and NGC\,2992.
These galaxies are strongly disturbed
 and can hardly be defined as disk galaxies anymore. 
In fact they are in-between disks and completely irregular galaxies.

Dividing the 716 galaxies with a normal sized (non-thick) bulge out of the RC3-survey
in asymmetric and symmetric in the same way as done for the
TBBs we find large differences between the samples. 
The fraction of asymmetric TBBs
is 54\,\% whereas the fraction of asymmetric TSBs is zero and of
asymmetric normal sized bulges 2.7\,\% (74\,\% of them are boxy and 26\,\% non-boxy) (Tab.~\ref{asy}). 
The latter values seem small compared to other studies but we use
a very strict definition of asymmetry  
and have neglected the galaxies with unclassifiable bulges in the RC3-survey due to
very strong perturbations/disruption.

\begin{table}[h]
\caption{Asymmetry of TBBs (RC3-survey)}
\begin{center}
\begin{tabular}{l||c|c|c|c}
 &  \multicolumn{4}{c}{g a l a x i e s \enspace w i t h \enspace a}\\
 &  &   &  normal sized &  normal sized \\
& TBB &TSB & boxy  & elliptical \\
&&& bulge & bulge \\
\hline \hline
asymmetric  & 7 & 0 & 14 & 5 \\
symmetric  & 6 & 5  & 303 & 394 \\
\end{tabular}
\end{center}
\label{asy}
\end{table}

\subsection{Individual notes on the asymmetric TBBs}
\begin{figure}[htb]
\centering
\includegraphics[angle=-90,width=9cm]{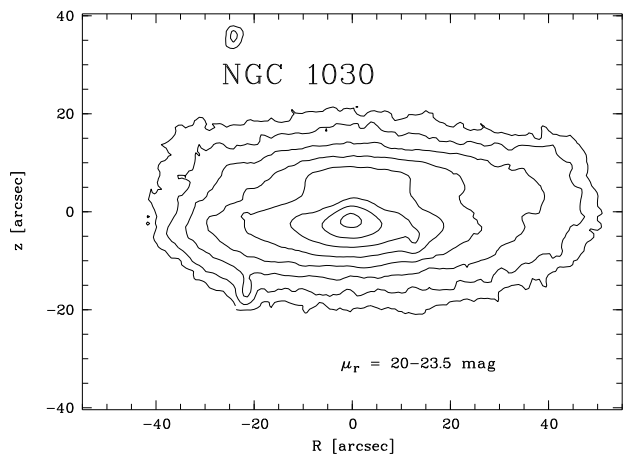}
\caption{NGC\,1030: Lowell/1.06m, 15min in $r$.}
\label{1030}
\end{figure}

In the following section we describe in detail the eleven prominent TBBs:

{\bf NGC\,1030} (Fig.~\ref{1030})
has a rectangular shape with merged bulge and disk
component  and the
asymmetries  grow with  distance from the major and minor axis.

{\bf NGC\,1055} (Fig.~\ref{1055})
is already well studied in the literature (e.g. Shaw \cite{sha93b}). 
The boxy isophotal distortions within the bulge component are spatially
extended and form almost a quadratic bulge. While on one side the disk is
bending away from the central plane and is
clearly separated from the bulge, on the other side bulge and disk overlap.
Additionally, the prominent dust lane on the northern side of the galaxy is
remarkable.

\begin{figure}[htb]
\centering
\caption{NGC\,1055: Contour plot from DSS.
JPG-image: http://www.astro.ruhr-uni-bochum.de/luett/tbb\_fig5.jpg}
\label{1055}
\end{figure}

The bulge of {\bf NGC\,1589} (Fig.~\ref{1589})
is merged with a thick asymmetric disk. The
boxiness is very pronounced in the inner part of the galaxy.

\begin{figure}[htb]
\centering
\caption{
NGC\,1589 has a prominent boxy bulge merging
into a thick disk. ESO/1.54m\,Danish, 20min in $I$.
JPG-image: http://www.astro.ruhr-uni-bochum.de/luett/tbb\_fig6.jpg}
\label{1589}
\end{figure}

The galaxy 
{\bf ESO\,383-005} (Fig.~\ref{38305})
--- previously studied by Kemp \& Meaburn (\cite{km})
pointing among other features also to
the considerable flat bulge ---
is the latest type (Sbc) in the sample of TBBs.
In the outer part of this galaxy 
its warped disk can be well separated from its large boxy bulge. 
Additionally, the disk is asymmetric in respect to its extent from the center
and is tilted with respect to the bulge ($\sim$\,3$^{\circ}$)
(Dettmar \& Barteldrees \cite{db90a}, L\"utticke \cite{lue96}).
One unusual substructure in
ESO\,383-005 
is remarkable (Fig.~\ref{38305}) and could be a possible merger remnant.

\begin{figure}[htb]
\centering
\caption{Latest morphological TBB with type (Sbc): The substructure in ESO\,383-005 at (R,z)\,=\,(38$\arcsec$,24$\arcsec$)
could be possibly a merger remnant. ESO/1.54m\,Danish, 20min in $r$.
JPG-image: http://www.astro.ruhr-uni-bochum.de/luett/tbb\_fig7.jpg}
\label{38305}
\end{figure}

The boxy bulge of {\bf NGC\,3573} (Fig.~\ref{3573})
is at low surface brightness levels
very asymmetric. On one side of the minor axis the bulge clearly ends and
a perturbed warped disk component follows, but on the other side bulge and
disk component are merged and largely spread into a diffuse structure which
is also asymmetric with respect to the major axis.

\begin{figure}[htb]
\centering
\caption{The asymmetric and strongly disturbed shape of NGC\,3573
at low surface brightness outer contours. ESO/1.54m\,Danish, 30min in $V$.
JPG-image: http://www.astro.ruhr-uni-bochum.de/luett/tbb\_fig8.jpg}
\label{3573}
\end{figure}

The most conspicuous component of {\bf ESO\,510-013} is its strong warped disk.
In the outer regions
this twisted disk contains
bright clouds of blue stars (Conselice \cite{con2001}).
Additional to a prominent dust lane aligned with the disk, there exists a second less
pronounced dust lane which is inclined by $\sim$\,8$^{\circ}$.
Their intersection is outside of the galaxy center.
As already mentioned in the ESO Press Information (\cite{EPI}) the prominent 
dust lane and the inner part of the bulge are not well aligned.
Especially the position angle 
of the main axis of the dust lane differ from
the main axis of the outer boxy part of the bulge. Since this extensive boxiness is
only visible at fainter brightness levels ($\mu > 21.6$ mag), 
it is possible that Conselice
(\cite{con2001}) has not noticed
this feature when he is describing ESO\,510-13
as galaxy containing ``a large spheroidal bulge''.
On the other side van Driel et al. (\cite{vd2000})
also mentioned the ``boxy structure''
of this galaxy.
A further remarkable feature of the bulge of ESO\,510-013 is
the excess of light on two diagonally opposite edges (possibly
marking the accreted satellite track and leading to the classification
``System possibly related to polar-ring galaxies'' (PRC
D-43, Whitmore et al. \cite{wh90}).

\begin{figure}[htb]
\centering
\caption{Contour plot of ESO\,510-13 from VLT. Besides the warped dusty disk
the enveloping boxy bulge with luminosity excess
on the lower-right and upper-left sides is remarkable: (R,z)\,=\,($-20\arcsec$,$50\arcsec$; $20\arcsec$,$-50\arcsec$).
ESO archive, 5min in $R$.
JPG-image: http://www.astro.ruhr-uni-bochum.de/luett/tbb\_fig9.jpg}
\label{510_13}
\end{figure}

\begin{figure}[htb]
\centering
\includegraphics[angle=-90,width=9cm]{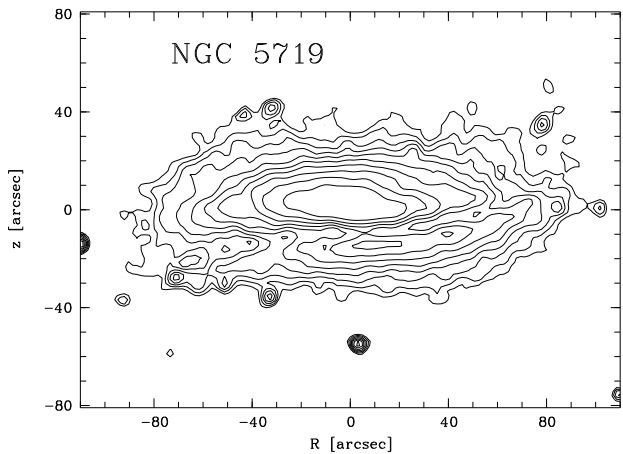}
\caption{NGC\,5719: Contour plot of NGC\,5719 from DSS.}
\label{5719}
\end{figure}

In {\bf NGC\,5719} (Fig.~\ref{5719}) the tilted dust lane is remarkable.
This dust lane is significantly bent and inclined to the major axis
of the galaxy.
The boxy structure in the outer parts of the bulge
envelops the inner dust distorted bulge.\\
Peletier \& Balcells (\cite{pb}) find  in NGC\,5719 differences in
$(B - R)$ and $(U - R)$ between bulge and disk of the order 0.2 mag
resp. 0.3 mag. These differences are nearly the largest values in their
sample of  30 galaxies, although the values are not extraordinary. 
Balcells \& Peletier (\cite{bp}) find a typical colour gradient within
the bulge reflecting the blueing
to the outer parts of the bulge.

\begin{figure}[htb]
\centering
\includegraphics[angle=-90,width=9cm]{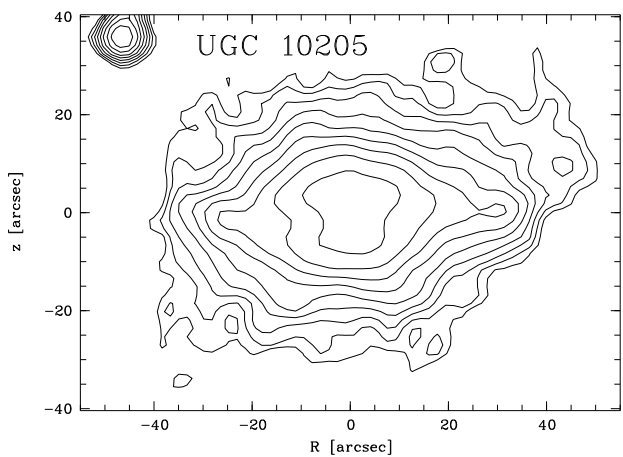}
\caption{UGC\,10205: Contour plot from DSS. The faint extended structure not visible in this image is to the lower left side of the galaxy.}
\label{10205}
\end{figure}

{\bf UGC\,10205} (VV 624) (Fig.~\ref{10205}) is already studied in detail by Reshetnikov \&
Evstigneeva (\cite{re99}) and Vega et al. (\cite{veg}).
Both note
that the bright spots in this galaxy (first mentioned by
Vorontsov-Velyaminov \cite{vv}) belong to an outer equatorial (edge-on) ring.
Other interpretations of this structure as a merger remnant (Vorontsov-Velyaminov \cite{vv})
or polar ring (Whitmore et al. \cite{wh90}, van Driel et al. \cite{vd2000}),
seem to  be very unlikely.
Reshetnikov \& Evstigneeva (\cite{re99}) point also to the asymmetric weak envelope surrounding the galaxy.
However, they do not mention its boxy structure at certain contour levels.
The asymmetry of the boxy envelope is especially remarkable
in the outer parts of the galaxy.
Additionally, Reshetnikov \& Evstigneeva (\cite{re99}) explain an ``extended, diffuse structure''
at one side of the galaxy,
only detectable in deep exposures (not visible in our DSS image, but in Fig.~2 in Reshetnikov \& Evstigneeva \cite{re99}),
as result of an accretion event.

{\bf IC\,4745} (Fig.~\ref{4745ccd}) is one of
the most prominent member of the galaxies with a
TBB and shows a lot of irregularities. Therefore it is surprising
that Bell \& Whitmore (\cite{bel}) do not mention its peculiar structure. 
The inner bulge system is asymmetric with respect
to the disk and the outer part of the bulge
(Dettmar \cite{det89}, Dettmar \& Barteldrees \cite{db90a})
as shown by a twist of
the isophotes visible in NIR images (L\"utticke \cite{luediss}).
The boxiness is most pronounced at lower surface brightness levels. 
Similar to NGC\,1055 the disk is on one side more extended 
and separated from the bulge, while
on the other side bulge and disk are merged at larger radial distances.
This impression is supported by the fact that the center of the boxy envelope
of the bulge is shifted ($\sim 5\arcsec$) to the
side of the merged disk/bulge
component.
Although many bright stars in the foreground
obscure the analysis of additional substructures,
one feature  at (R,z)\,=\,($-40\arcsec$,$25\arcsec$) is notable (Fig.~\ref{4745ccd}).
It can be associated with a possible merger remnant.
Perhaps related to this structure is a further feature which is in the outer
region of the galaxy
at (R,z)\,=\,($-55\arcsec$,$35\arcsec$).
A dissolving satellite could be an explanation for this structure.

\begin{figure}[htb]
\centering
\caption{IC\,4745 is the prototype of the galaxies with a TBB because of its many
irregularities, substructures, and asymmetries.
ESO/1.54m\,Danish, 20min in $R$.
JPG-image: http://www.astro.ruhr-uni-bochum.de/luett/tbb\_fig12.jpg}
\label{4745ccd}
\end{figure}

\begin{figure}[htb]
\centering
\includegraphics[angle=-90,width=9cm]{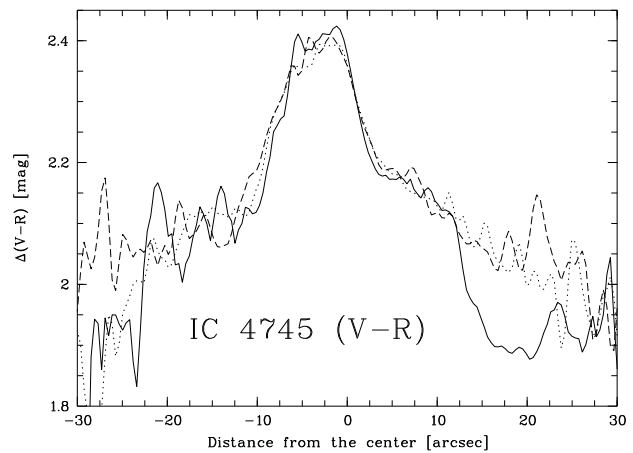}
\caption{
In cuts of the colour map through the bulge 
the blue gradient is eminent.
Minor axis cut: solid line.
Cut with an angle of 60$^{\circ}$ to the major axis: dotted line.
Cut with an angle of $-60^{\circ}$: dashed line.
}
\label{4745colourcuts}
\end{figure}

The blueing with
radial distance is evident in the $(V\!-\!R)$ colour profiles (Fig.~\ref{4745colourcuts}).

{\bf IC\,4757} exhibits a boxy bulge extending into a boxy envelope
on all contour levels, 
without showing a normal disk component (Fig.~\ref{4757}).

\begin{figure}[htb]
\centering
\caption{Deep image of IC\,4757: NTT/ESO, 30min in $R$.
JPG-image: http://www.astro.ruhr-uni-bochum.de/luett/tbb\_fig14.jpg}
\label{4757}
\end{figure}

\begin{figure}[htb]
\centering
\includegraphics[angle=-90,width=9cm]{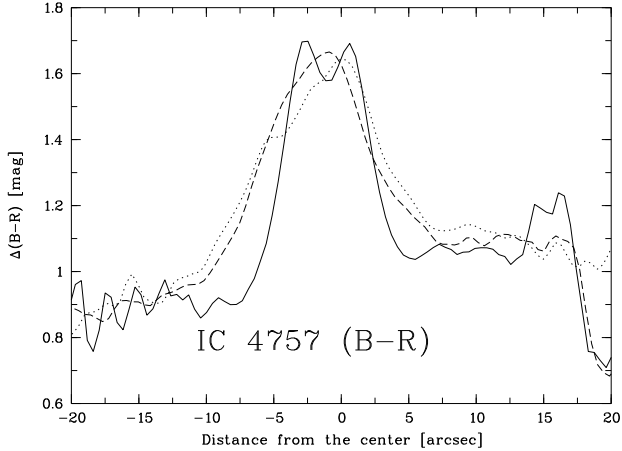}
\caption{
Color profiles in $(B\!-\!R)$
reveal  the asymmetry in the bulge. 
The colour above the plane (positive values of the abscissa) is
obviously redder than the colour below the plane.
Minor axis cut: solid line.
Cut with an angle of 45$^{\circ}$ to the major axis: dotted line.
Cut with an angle of $-45^{\circ}$ to the major axis: dashed line.}
\label{4757colourcuts}
\end{figure}

While its asymmetry in the optical image is difficult to see, it is noticeable
in the colour $(B\!-\!R)$.
The colour difference between the parts of the bulge
above and below the dust lane, $\Delta(B\!-\!R)$\,=\,0.2\,mag, is larger than
the colour differences in each part,
$\Delta(B\!-\!R)$\,=\,0.1\,mag (Fig.~\ref{4757colourcuts}).
This unusual result can
not be explained by dust reddening because the red side of the bulge is
not affected by the dust lane.  Bad alignment between
the images of the different filters
can also be excluded due to the lack of asymmetric residuals of stars in
the colour map (cf. L\"utticke \cite{luediss}).

\begin{figure}[htb]
\centering
\includegraphics[angle=-90,width=9cm]{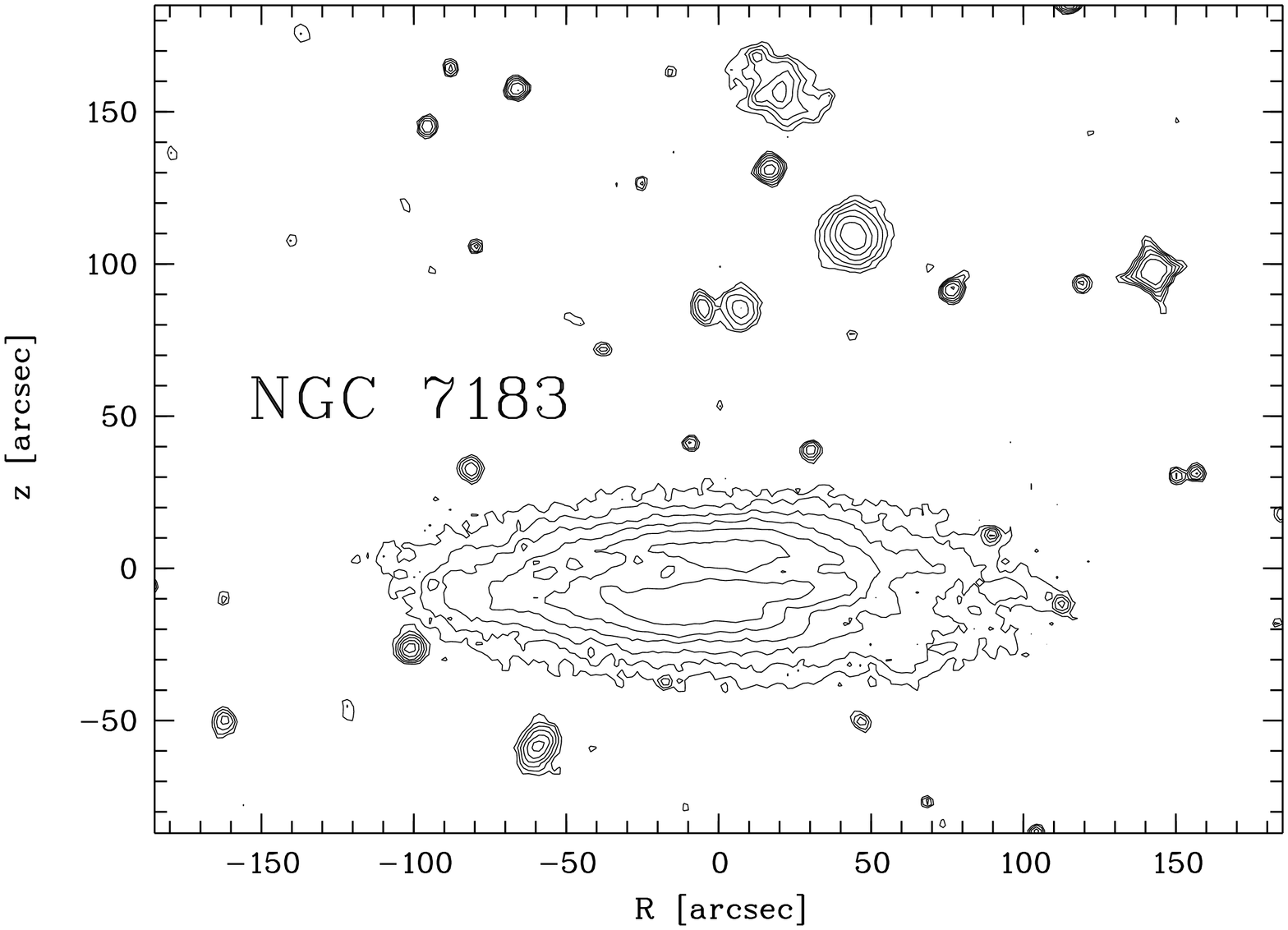}
\caption{NGC\,7183 and its projected satellite system: 
(R,z)\,=\,($-60\arcsec$,$-60\arcsec$), ($-5\arcsec$,$85\arcsec$), ($5\arcsec$,$85\arcsec$), ($20\arcsec$,$160\arcsec$), ($110\arcsec$,$45\arcsec$).
Contour plot from DSS.}
\label{7183}
\end{figure}

{\bf NGC\,7183} (Fig.~\ref{7183}) has a bulge with a  disturbed boxy shape and
notable asymmetries. In the outer parts the merged disk/bulge component is
asymmetrically extended.

\subsection{Luminosity profiles}
Studies in the literature reveal that the luminosity distribution of bulges
is well fitted by the Sersic $r^{1/n}$-law. 
The
shape parameter $n$ is correlated with the Hubble type in the sense that
late type galaxies have exponential bulges ($n$\,=\,1) and bulges of
early types are closer to the
original de Vaucouleurs-law (\cite{vau})
with $n$\,=\,4 (Andredakis et al. \cite{and},
Courteau et al. \cite{cou}).

\begin{figure}
\centering
\includegraphics[angle=-90,width=9cm]{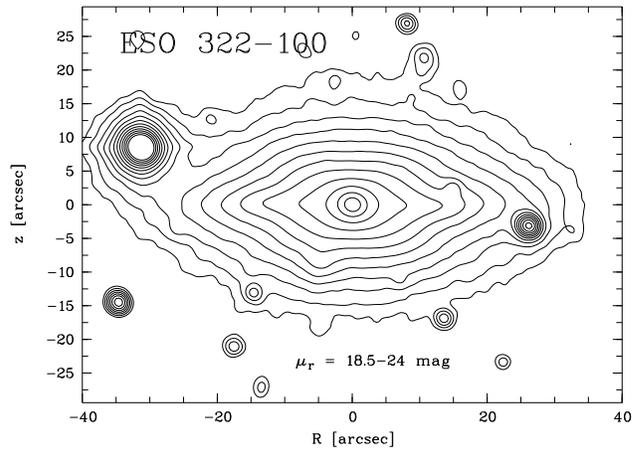}
\caption{
ESO\,322-100: ESO/2.2m, 15min in $r$.}
\label{322100}
\end{figure}

\begin{figure}
\centering
\includegraphics[angle=-90,width=9cm]{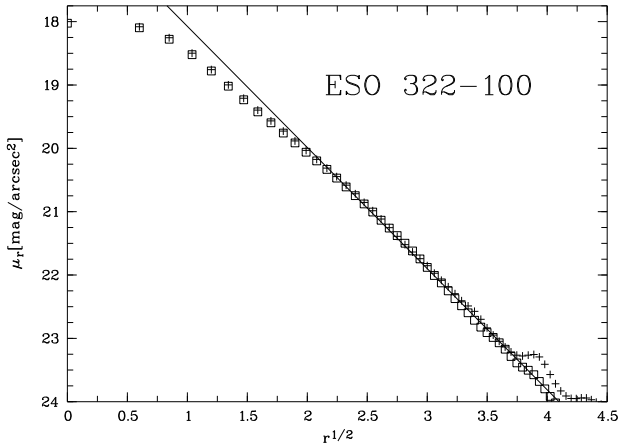}
\caption{
The TBB of the S0 galaxy ESO\,322-100
is described best by the shape parameter $n$ = 2.
This reveals the luminosity profile along the minor axis in the outer bulge parts.
The surface brightness approximately correlates with $r^{1/2}$.
The cut above the  plane is marked by crosses, below the plane by squares.}
\label{n_cuts}
\end{figure}

The often asymmetric and merged bulge and disk components severely hampers
any bulge/disk decomposition.
Therefore it is only possible to 
determine the shape of the luminosity distribution
of the bulge along the minor axis
to minimize the influence of the disk in edge-on galaxies.
The analysis of the
luminosity profiles, testing only $n$\,=\,1, 2, 3 and  4,
shows that $n$\,=\,4 is only
for one (UGC\,10205) out of thirteen investigated TBBs the best description.
In this galaxy, the minor axis surface brightness plotted
against $r^{1/4}$ shows two components.
Their intersection is at z\,=\,$\pm 10\arcsec$. Three bulges
have a shape parameter of $n$\,=\,3, while seven bulges,
even in early type galaxies (Fig.~\ref{322100} and \ref{n_cuts}), follow the
$r^{1/2}$-law. Two galaxies have
even exponential bulges (Tab.~\ref{cat}).
There is no clear correlation for the galaxies with a TBB between $n$ 
and the Hubble type.

\subsection{Bar detection}
Kinematical studies (Vega et al. \cite{veg},
Merrifield \& Kuijken \cite{mer}, Garcia-Burillo \cite{gb})
and bar detection in radial cuts of NIR images (L\"utticke \cite{luediss}), 
Vergani et al. \cite{vpld}, Paper II)
reveal that the most galaxies with a TBB do not habour a bar.
In detail, from nine TBBs studied only one possesses a bar, whereas six do not
show a bar signature (Tab.~\ref{nobar_bar}).
For two galaxies there are controversial results.
Garcia-Burillo (\cite{gb}) and Vergani et al. (\cite{vpld}) detect a bar
in NGC\,1055, Merrifield \& Kuijken (\cite{mer}) did not.
UGC\,10205 is denoted as unbarred by Reshetnikov (\cite{res03}) and
L\"utticke (\cite{luediss}) while Vega et al. (\cite{veg})
propose that the galaxy harbours a bar.
However, inspecting their data  the interpretation of the
velocity curve of UGC\,10205 by a possible bar
seems to be unlikely (cf.~Sect.~4.1.2).

\begin{table}[h]
\caption{Bar detection in TBB galaxies}
\begin{center}
\begin{tabular}{l||c|l|l}
Galaxy &  Bar & \multicolumn{2}{c}{Reference$^1$}  \\
& & for a bar &  for no bar \\
\hline \hline
NGC\,1030 & no & & V01 \\
NGC\,1055 & ?  & G01, V01 & M99 \\
NGC\,1589 & yes & V01 & \\
UGC\,3458 & no & & L99 \\
ESO\,506-003 & no & & L99 \\
UGC\,9759 & no & & L99, V01 \\
UGC\,10205 & ?$^2$ & V97 & L99, R03 \\
IC\,4745 & no & & L99 \\
NGC\,7183  & no &  & V01 \\
\end{tabular}
\end{center}

$^1$: G01: Garcia-Burillo (\cite{gb}), L99: L\"utticke (\cite{luediss}),
M99: Merrifield \& Kuijken (\cite{mer}), R03: Reshetnikov (\cite{res03}),
V97: Vega et al. (\cite{veg}), V01: Vergani et al. (\cite{vpld})\\
\\
$^2$: Most likely: no (cf.~text). \\
\label{nobar_bar}
\end{table}

\subsection{Environment}
To investigate environmental effects on the
galaxies with a TBB, we have examined the large
scale as well as the small scale environment.

3033 out of the total sample of 6392 galaxies (47\,\%) in Garcia (\cite{gar})
belong to groups, but the fraction of galaxies with a TBB
in groups is significantly larger (77\,\%, Tab.~\ref{cat}).  However, a direct
comparison with a control sample 
of non-b/p shaped bulges (drawn from L\"utticke \cite{luediss})
with similar Hubble types (S0 - Sb)
also exhibits a fraction of 67\,\% of
galaxies in groups.
This reveals that the rate of galaxies with a TBB in groups is not
significantly increased.

In fields around the center of a galaxy with a TBB 
scanning a radius
of $5 \times D_{25}$ resp. $2.5 \times D_{25}$ there is nearly
 no difference in the number distribution
of confirmed companions ($\Delta v$\,$<$\,1000\,km\,s$^{-1}$) with known velocity (derived from NED\footnote{NASA Extragalactic Database})
compared to the control sample. E.g.,
63 \% of the galaxies with a TBB have no
companion inside $5 \times D_{25}$  and
the mean value of companions is $0.6 \pm 0.2$, while in the control sample
the fraction is 68\,\% with the same mean value.

However, inside $1 \times D_{25}$ around a galaxy with a TBB
84\,\% of these galaxies have
at least one projected satellite visible on DSS images.
The control sample only shows a fraction of
32\,\%. This dependence between the existence of nearby satellites and
TBB is statistically significant on a 0.1\,\%-level ($\chi ^2$-test) 
(Tab.~\ref{sat}).

The nearby satellite system of NGC\,7183 with at least five projected galaxies
is especially remarkable (Fig.~\ref{7183}). IC\,4745 has also
a prominent  nearby
satellite system including at least three galaxies.
However, the satellite
system has a larger extent 
than the system of NGC\,7183.
Furthermore, NGC\,3573 has one projected satellite
only $2\arcmin$  and
one confirmed companion (ESO\,377-017, $\Delta v$\,=\,22\,km\,s$^{-1}$) $12\arcmin$ 
away.

\begin{table}[h]
\caption{Number of galaxies with projected satellites}
\begin{center}
\begin{tabular}{l||c|c}
 &  galaxies with & galaxies without  \\
 &   a TBB & a b/p bulge \\
\hline \hline
no projected satellite  & 5 & 21  \\
projected satellite(s)  & 16 & 10  \\
\end{tabular}
\end{center}

\begin{small}
A galaxy is here defined as projected satellite, if it is inside a radius of the
primary galaxy diameter
($1 \times D_{25}$) around this galaxy.
\end{small}
\label{sat}
\end{table}

Determining the normalized
interaction index $I$, defined by van den Bergh et al. (\cite{vB}),
the galaxies with a TBB
reveal a large value of $1.7 \pm 0.2$ (Tab.~\ref{ii}) which is even larger than
the interaction index of 
the Hubble Deep Field ($I_{HDF}$\,=\,1.0, van den Bergh et al. \cite{vB}).
Regarding the morphological  pecularities of the TBBs such a
large interaction index 
for this group of galaxies is not surprising,
but it quantifies the characteristic of these galaxies.

\begin{table}[h]
\caption{Interaction Index}
\label{ii}
\begin{center}
\begin{tabular}{l|lccccc|c}
sample & $w$: & 0 & 1 & 2 & 3 & 4 & $I$ \\
\hline
galaxies with & & 0 &  7 & 10 &  2 & 0 & $1.7 \pm 0.2$ \\
a TBB & & & & & & & \\
\hline
galaxies without & & 26 & 3 & 2 & 0 & 0 & $0.2 \pm 0.1$ \\
a b/p bulge & & & & & & & \\
\end{tabular}
\end{center}

with $w=0$: objects with no tidal distortion, 
$w=1$: objects with possible tidal distortion, 
$w=2$: objects with probable tidal effects,
$w=3$: possible merger, $w=4$: almost certain merger
\end{table}

\section{Discussion}
\subsection{The origin of the box structure in the TBB}
\subsubsection{Bars ?}
Results of different studies (cf.~Sect.~3.4)
show that more than 60\,\% of the galaxies with
a TBB have no bar.
Therefore not all TBBs can be explained by the existence of bars.
However,
there is the possibility that former bars (which are now dissolved) could
be responsible for the boxy structure of the TBBs.
It is known that bars can be destroyed by large gas inflow and mass
accumulation in the center (Hasan et al. \cite{hpn93}, Friedli \& Benz
\cite{fb93}, Norman et al. \cite{nsh96}, Sellwood \& Moore \cite{sm99},
Bournaud \& Combes \cite{bc02}, Berentzen et al. \cite{bahf03})
and boxy bulges arise progressively at the end of the bar
life-time (Combes \cite{com03}).
On the other side, recent studies show that bars are more robust
against central mass concentration than previously thought
(Shen \& Sellwood \cite{ss03}).

Is it therefore possible that all TBBs have their origin in dissolved
or existing bars ?

In general, TBBs are often more box shaped in the outer
than in inner part of the bulge in contrast to normal b/p shaped bulges.
Regarding simulations the shape of boxy bulges produced by a bar look very different to the observed TBBs in vertical as well as in radial extent 
(cf. Combes et al. \cite{com90}, Paper II, Athanassola \& Misiriotis \cite{am02}, Patsis et al. \cite{psa02}).
Quanfifying this result it is evident that bars can only be responsible
for boxy structures smaller than $D_{bar}$ and for boxy structures with
larger vertical extent smaller than 0.7\,$\times$\,$D_{bar}$
(e.g. Athanassola \& Misiriotis \cite{am02}, Patsis et al. \cite{psa02}).

Measurements for samples of strong bars give medium relative bar lengths
of $D_{bar}/D_{25} = 0.37 \pm 0.03$ (Sab-Sc galaxies, derived from
Elmegreen \& Elmegreen \cite{ee95}) resp. $0.42 \pm 0.05$ (S0-Sb galaxies,
derived from Paper II setting BAL\,=\,$D_{bar}$ in galaxies with a
peanut-shaped bulge).
In the literature there are only a few measurements detecting bar
lenghts ($D_{bar}/D_{25}$) larger than 0.5 (Elmegreen \& Elmegreen
\cite{ee85}, Otha et al. \cite{oth90}, Elmegreen \& Elmegreen \cite{ee95},
Shaw et al. \cite{sha95}, Paper II, Erwin \& Sparke \cite{es03}).
Using a classical method (cf. Erwin \& Sparke \cite{es03}) Otha et al. (\cite{oth90})
find the only bar (as far as we know)
with a relative length larger than 0.6.
Using a new method to quantify the length of a bar, Erwin \& Sparke
(\cite{es03}) find additional four bars with values larger than 0.6.

How about the origin of the TBB for galaxies with a bar detection ?
In the case of NGC\,1055 (cf.~Sect.~3.4)
the position of the end of the (possible) bar, derived from
radial NIR profiles, is in the range of the turn-over point of
the rotation curve at a radial distance of R\,=\,$30\arcsec$
(measured at a vertical distance of z\,=\,$-15\arcsec$
below the major axis, well out of the dust lane) (Shaw \cite{sha93b}).
Since the box shape of the bulge is most pronounced at R\,=\,$120\arcsec$
(cf.~Fig.~\ref{1055}), the existing bar potential cannot explain this box
structure.
However, another possibility would be to assume that there was
a larger (now dissolved) bar present in this galaxy. This could be
explained
by the evolutionary scenario with bar formation and renewal of
Bournaud \& Combes (\cite{bc02}). They show in simulations that former bars are
relative larger than later ones.
However, the differences of the lenghts of the bars in their simulations
are far away from a factor of four, which we would need in the case of
NGC\,1055 to conclude from an existing bar of a projected length of
R\,=\,$30\arcsec$ to a former bar of R\,=\,$120\arcsec$.

In addition, it is not yet clear how shorter bar lenghts after the bar
renewal processes fit to an evolution along the Hubble sequence from
late- to early-types, since all observations
reveal that bars in late-type galaxies are shorter with respect to the
galaxy size compared to early-type galaxy bars (Elmegreen et al.
\cite{el96},
Erwin \& Sparke \cite{es03}, Erwin \cite{erw03}).

To conclude: The small number of observed bars in TBBs, the boxiness of
TBBs in their outer parts in comparison to normal b/p bulges, the size
of the TBBs in relation to observed bar lenghts, and the lack of any
explanation for the large TBB of NGC\,1055 by a present or dissolved
bar are strong indications against a scenario of TBBs originating
by bars.

There is of course the possibility that in some galaxies with a TBB the
inner boxy structures could be produced by the bar in the usual way but
for the outer boxiness there have to exist another explanation.

Therefore it is not possible that the galaxies with a TBB are only the
most early types in the normal b/p category which is in general associated with bars.
We have found a lot of S0 galaxies with a b/p bulge which have small bulges
(that means BUL/D$_{25} < 0.5$) and do not belong to the class of TBB galaxies (Paper I).
In addition, the ratios of the luminosity of the bulges to the total
luminosity of galaxies with normal b/p bulges 
are in the most cases significally larger compared to the TBB galaxies.
This is the reason why TBB galaxies are classified
in the range S0 to Sbc, although their bulges have a large spatial size. 
However, size
is not a criterion for classification but luminosity.

\subsubsection{Soft mergers ?}
Regarding the merger/accretion scenario for the origin of b/p bulges 
(cf.~Sect.~1)
Vergani et al. (\cite{vdk}) find evidences for
a soft merger between NGC\,1055 and a satellite analysing 21-cm HI observations.
Asymmetries in the distribution and in
the kinematics of the neutral hydrogen and particularly
the asymmetric clumpy HI emission in the outermost
northwestern region indicate an unrelaxed structure
typical for tidal interactions and accreted material enlarging the bulge.
In further three investigated TBBs the HI is also distributed in a
peculiar way. One galaxy has a strong warped disk (ESO\,383-005), 
the other harbours significant non-settled
(perhaps counter-rotating) gas (IC\,4745), and the third  exhibits a patchy
 HI distribution in general (IC\,4757) (Vergani et al. \cite{vdkag}).

The two components of the bulge of UGC\,10205 detected in the
analysis of the luminosity distribution in UGC\,10205 (cf.~Sect.~3.3)
are in agreement with 
the multi-component structure of the gas revealed by
Vega et al. (\cite{veg}). They disentangle three kinematically distinct
gaseous components for this galaxy.
Two of them have quite similar velocity dispersion profiles, but very different
velocity curves. Vega et al. (\cite{veg})
interpret them by a possible bar due to
a similarity to velocity curves with  a ``figure-of-eight'' appearance
which is the signature of  a bar
(Kuijken \& Merrifield \cite{kui}).
They associate  the third component to the faint features embedding the galaxy. 
However, the ``figure-of-eight'' signature is 
only very weak and  Reshetnikov \& Evstigneeva (\cite{re99}) suggest a
different origin:
One gas component is determined by the rotation of the bulge, 
another by the overall potential of the bulge and disk, and the peculiar inner
subsystem
is the result of accreted material from a small companion.
They support this  conclusion  by the observation that 
the star-formation rate is fairly high compared to normal disk galaxies.
In addition,
they  run a numerical simulation of a capture and tidal disruption of a satellite 
galaxy by UGC\,10205 producing the observed asymmetric envelope structure (Reshetnikov \& Evstigneeva \cite{re99}).

The evolutionary scenario in which 
TBBs are formed by soft mergers is also probable for
the prototypical case, IC\,4745.
This galaxy shows many features supporting an accretion scenario:
substructures which could be possible merger
remnants or merging satellites, a  numerous satellite system pointing to
a high density of galaxies, a significantly asymmetric envelope, 
a colour gradient becoming bluer at increasing
radial distance, and even asymmetries
reflected in the rotation curve
on the major axis (Dettmar \& Barteldrees \cite{db90a}).
All these observations lead to the conclusion that the bulge structure
of IC\,4745  has been
formed just recently and is not yet dynamically settled  similar to UGC\,10205. 

All other galaxies of this class
also show at least some indications (cf. Sect.~3.1) 
for a merger induced origin of the TBBs
which is further supported by the generally stronger asymmetries of these galaxies 
compared to the galaxies without a TBB (Tab.~\ref{asy}).

The soft merging scenario is able to explain the 
deformed structures of asymmetric or thick disks and
the merged bulge and disk components which are not produced in 
a bar scenario. 
The strongly warped dusty disks of the galaxies with a TBB
are indications for such recent interactions
(Conselice \cite{con2001}) and
the observed tilted disks could also point to the effects of
the gravitational forces (however, triaxiality of bulges
is also a possible explanation for tilted disks).
Furthermore, unusual components and structures
in addition to disk and bulge
(as e.g. in UGC\,10205) and high star formation rates
(UGC\,10205, Reshetnikov \& Evstigneeva \cite{re99}; ESO\,510-014, Conselice \cite{con2001})
are evidences
for a recent merger event.
Since such events often trigger bars (Noguchi \cite{nog}, Gerin et al. \cite{ger}, Walker et al. \cite{wmh})
there is a combination of bars and mergers building TBBs,
but the bars do not seem to be responsible for the extended boxy structures 
(cf. previous section).  

The increased fraction of small
nearby projected satellites coupled with a large interaction index
also supports the conclusion that the environment
of galaxies with a TBB has an increased density of small
galaxies and former or ongoing accretions of
satellites seem to be very likely.

\subsubsection{Alternative scenarios ?}
Explaining the  box structure only by interactions,
as shown by May et al. (\cite{may}),
external cylindrically symmetric torques are required.
However, these torques arise only in very special conditions.
Therefore
the fraction of such bulges produced in this way is very small or such
 bulges do not exist at all.
Perhaps the boxy bulge of NGC\,3573
marked by large distortions and
the nearby satellite and companion
could be a result of such a special
interaction event.

Using N-body simulations Patsis et al. (\cite{pags}) 
have shown that boxy edge-on
profiles can also be accounted in models of normal
non-barred galaxies by the presence
of vertical resonances populating stable families of
periodic orbits.
However, their simulations reveal that the size of the boxy structures are
again too small in comparison to the box-size of the TBB galaxies.
Therefore this model is not applicable for the TBBs.

\subsection{Merger scenario: theory and observation}
The origin of the TBBs by accretion events can be
theoretically explained by the study of
Binney \& Petrou (\cite{bin}). The special conditions needed
for an accretion leading to a boxy bulge in their investigation correspond
to the small number of TBBs. Only 3.9\,\%  of 330
galaxies with a b/p bulge  
or 1.8\,\% of 734 disk galaxies 
derived in the RC3  survey 
have a TBB.

In the accretion scenario by Whitmore \& Bell (\cite{whi})
a b/p bulge arises if the impact of the satellite is
at an oblique angle.
Dynamical friction begins to shred the companion on an inclined encounter.
The gas settles into a new
oblique disk and cloud-cloud collisions
therein cause rapid star formation.
While the individual resulting stars still relatively circularly orbit around
the center of the galaxy, the oblique disk spreads out into a precession cone.
In projection this cone will appear as an X-shaped structure which superposes
the bulge forming the apparent b/p shape.
Hernquist \& Quinn (\cite{her})
support by their N-body simulations the possibility of
the origin of an X-shaped structure by an accretion event. 
However, the
added mass is small in comparison to the disk in their simulation, 
while Whitmore \& Bell (\cite{whi})
suggest that for IC\,4767 (a galaxy with a normal b/p bulge) 
the mass in the X-component has the same size
as the mass in the disk.
 Even taken into account that the X-component has its origin not only in accreted material,
but also in recently formed stars, the accreted companion must have a large
mass in the picture of Whitmore \& Bell (\cite{whi}).
From the theoretical point of view it is likely that any
axisymmetry is destroyed (Combes et al. \cite{com90})
and boxiness arises also in outer parts
of a galaxy (Schweizer \& Seizer \cite{sch}) after the merging
of such a large amount of mass rotating at a high speed (Combes et al. \cite{com90}).
Therefore an accretion scenario is not likely for galaxies
with normal b/p bulges which are very symmetric and harbour the b/p structure
in the inner part of the galaxy. 

However, the scenario of Whitmore \& Bell (\cite{whi})
is a possibility for the formation of the TBBs
because they show the expected asymmetry after a merger event and
the boxiness is often most pronounced at fainter
surface brightness levels leading to the fact that accreted
low mass satellites can produce
the observed boxiness.  
Evidence for such an evolutionary picture can be found
in  ESO\,383-005, IC\,4745, ESO\,510-013, and UGC\,10205.  
The first three galaxies have
possible merger remnants at one edge of the box respectively  
 a luminosity excess at two opposite edges. 
IC\,4745 shows an additional indication for a former accretion event,
namely the predicted  X-structure which is visible in $(V\!-\!R)$
(L\"utticke \cite{luediss}).
 UGC\,10205 seems to capture material of a disrupted satellite whose
 remaining structures are still visible.

\subsection{Merger scenario: consistency check}
 As a consistency check for our merger scenario we can compare the observed number of
TBB galaxies in our RC3-survey with a roughly calculated number of 
TBB galaxies expected from a general accretion rate of satellites.
At first we have to answer the question which satellites 
can produce a TBB.
We restrict the parameters for our calculation to the two most important ones, namely mass and impact angle.
Measuring the angle $\theta$ between the major axis and the ray passing through
the edge of the boxy bulge structure 
(Shaw et al. \cite{sh90}, L\"utticke \cite{luediss})
and associating this angle with the infall parameter of the satellite orbit
we get an estimate for the range of oblique impact angles which are needed 
to produce a boxy structure (Whitmore \& Bell \cite{whi}).
We find angles in the range of $\theta$ = [35$^0$,53$^0$], 
i.e. $\sim$20\,\% of all angles are suitable 
if we suppose that all impact angles are equiprobable (Zaritsky \& Gonzalez \cite{zg99}).
To find out which range of satellite masses can produce a TBB we have looked into 
several simulations in the
literature.
Walker et al. (\cite{wmh}), Huang \& Carlberg (\cite{hc97}),
Velazquez \& White (\cite{vw99}), and
Font et al. (\cite{fnsq})
analyse minor mergers in which the satellite has 1-4\,\% of the mass of
the parent galaxy (i.e. 10-30\,\% of the disk mass).
The results are disk thickening, warps, small growth of the bulge etc., but
no large scale change of the morphology of the bulges. 
On the other side of the mass scale Barnes (\cite{bar99}) and Bendo \& Barnes (\cite{bb00})
investigate medium-sized mergers with ratio 1:3 between
two disk galaxies.
The remnant is described as ellipsoidal with fairly regular disk-like kinematics.
Since the TBB galaxies are still disk galaxies such mergers are too
violent. 
Reshetnikov \& Evstigneeva (\cite{re99}) estimate from their observational data
that the mass of the merger
remnant of the satellite of UGC\,10205 is 4\,\% of the parent galaxy, while
the initial mass of the satellite is $\sim$10\,\% (Reshetnikov \cite{res03}).
Regarding these studies we suppose that the suitable mass ratio between
the mass of satellite and parent galaxy for the generation of a TBB has
to be larger than 1:20 and smaller than 1:3. 
Therefore an appropriate mass fraction of
the satellite seems to be $\sim$10-15\,\% (i.e. $\sim$1:10-1:6).
Zaritsky \& Rix (\cite{zr97}) propose for such kinds of mergers (satellite has
$\sim$10\,\% of the mass of the parent galaxy) an upper satellite accretion rate of
0.07-0.25 per Gyr for local spiral galaxies. 
The large uncertainty in this upper limit is explained in Zaritsky \& Rix (\cite{zr97}).
Using this accretion rate as an upper limit for the calculation of the numbers of TBBs in our RC3-survey
we get: \\
734 (galaxies in the survey) $\times$ 0.07-0.25 per Gyr $\times$
0.2 (fraction of satellites with
suitable impact angle) = 10-37 TBB galaxies per Gyr.

If we further assume that the precession cone building the boxy structure arises
in a late phase of a suitable
soft merger (cf. Fig.~3 of Hernquist \& Quinn \cite{hq}),
i.e. in the last quarter of this merger which will take about one Gyr (Balcells \cite{bal}),
and that the asymmetries as signature for an accretion are visible $\sim$1-2 Gyr
(Zaritsky \& Rix \cite{zr97}, Reshetnikov \& Evstigneeva \cite{re99}, Balcells \cite{bal})
we estimate that the asymmetric TBB galaxies have a lifetime of $\sim$2 Gyr.
Therefore we get in summary an upper limit for our RC3-survey of
20-74 asymmetric TBB galaxies.

Compared to the seven detected asymmetric TBB galaxies our observations
are below this upper limit.
A calculation of the upper limit for the number of all (asymmetric and symmetric) TBB galaxies
is not possible since we do not yet know how long their thick boxy structure exists,
i.e. if there are other mechanisms, such as subsequent mergers or interactions,
which can destroy or smooth a TBB.

\subsection{TBB as a stage of secular evolution}
The  luminosity distributions of the TBBs
are marked by the lack of
any correlation between the shape
parameter $n$ and the Hubble type and
by the often best description of the bulges by a $r^{1/2}$-law,
also in early type galaxies.
An interpretation for the
non-existing correlation is that the TBBs could be involved into
recent formation processes in which the Hubble type
is changing.
This scenario is supported by studies revealing that an accretion of a satellite
leads to a growing bulge and to an increasing  profile index $n$
proportional to the satellite mass (Andredakis et al. \cite{and},
Aguerri et al. \cite{agu}).
Additionally, such an interpretation is in accordance with
the unusual prominent dust lanes
in 18 of the 19 TBB galaxies (even in S0-galaxies),
since normally such dust lanes are typical
for later types.
Therefore we conclude that TBB galaxies are in-between two positions in the
Hubble sequence.
We suppose that galaxies with a TBB could be in a process
in which the Hubble type is changing ($\Delta T$\,=\,2-4).
In this way theories of secular evolution including transformation
of later Hubble types into earlier (e.g. Pfenniger \cite{pfe2000})
get observational support.

Regarding the asymmetric TBBs  and
the TSBs  it is possible
that the boxiness depends on the aspect angle.
Another, additional explanation for the degree of boxiness is given by
Schweizer \& Seizer (\cite{sch}). Their results reveal that boxiness
could be related to the time of the last merger event. This is
supported by the fact that the boxiness of the rather symmetric
 TBBs is
less pronounced in the outer bulge part than
in the asymmetric TBBs.
In  such a scenario the TBBs will become dynamically settled,
the bulges develop into spheroidally shaped and symmetries,
and relicts of the merger
are then only features at very low surface brightness.
This picture gets further support by the detection
of such extended features  at $\sim$\,28\,B-mag$\cdot$arcsec$^{-2}$
in ``normal'' early type
disk or TSB galaxies (e.g. Sombrero Galaxy)
(Malin \& Hadley \cite{mar99}).

\subsection{Summary}
In a merger sequence arranged by the effects of a merger  on the
central component of the parent disk galaxy 
the mergers resulting in TBBs can be integrated as follows:
The smallest
mergers have almost no effect (possibly small growth of the bulge) 
or let arise bars triggering box- or peanut shaped
bulges, stronger mergers with an oblique impact angle result in TBBs,
large mergers in elliptical galaxies,
and the heaviest collisions destroy galaxies.
Thus galaxies with a TBB  stand between ``normal'' disk
or lenticular galaxies and elliptical galaxies with small disks
in respect of the mass of the accreted companion.
Regarding the small fraction of TBBs we suppose that mergers
resulting in TBBs are only
a  secondary process for the secular evolution of disk galaxies.

\section{Conclusion}
We present and characterise a new class of disk galaxies. 
They are defined by:
\begin{itemize}
\item unusually large bulges: BUL/$D_{25} > 0.5$
\item box-shaped bulges: bulge type 3 or 2 (Paper I)
\end{itemize}
We call  the bulges of these galaxies
``Thick Boxy Bulges'' (TBBs) according to their appearance.
Using DSS, CCD, and NIR data we  derive the following results:
\begin{itemize}
\item 2\,\% of all disk galaxies (S0-Sd) belong to the class of TBB galaxies.
\item 4\,\% of all galaxies with a b/p bulge have a TBB.
\item The extent of the box shape in TBBs seems to be too large to result
from a  bar potential.
\item In general the morphology of the TBB galaxies is disturbed and
a large fraction of these galaxies is very asymmetric.
\item There exist two kinds of b/p bulges. The ``normal''
b/p bulges (96\,\%) are triggered by bars, while the rest -- the TBBs --
have most likely their origin in accretion of satellites (soft mergers).
Bars can also exist in TBB galaxies, or have been present and are now dissolved, however, they seem to be
only responsible for the inner and not for the extended boxy structures of TBBs.
\end{itemize}

The merging scenario is especially
supported by many morphological peculiarities (e.g.\,merger remnants) of the
TBB galaxies.
Additionally, we propose a possible ongoing 
change of the Hubble type in galaxies with a TBB.
Our studies of colours, environment, luminosity and dust distribution and
recent studies of kinematics and HI (Vergani et al. \cite{vdk} and \cite{vdkag}) give further
indications for
a merger scenario.
More evidences for such a scenario come from the comparison between theories 
of soft merging and our observations,
the detailed analysis of the TBB galaxy UGC\,10205 (Reshetnikov \& Evstigneeva \cite{re99}),
and the detection of high star formation rates in two TBB galaxies.
The number of observed TBB galaxies is consistent with an
upper limit derived from general satellite accretion rates. 
However, further investigations about star formation history and simulations of minor mergers
are needed for a complete understanding of
the formation process  and the lifetime of TBBs.

\begin{acknowledgements}
Part of this work was supported by the
\emph{Deut\-sche For\-schungs\-ge\-mein\-schaft, DFG\/}.
This work is partly based on observations obtained at ESO/La Silla,
DSAZ/Calar Alto, and Lowell Observatory.
Furthermore, this research has made use of the VLT Science Archive
(used observations were made with ESO Telescopes at the Paranal
Observatory under programme ID 60.A-9203(A).)
and the NASA/IPAC Extragalactic Database
(NED) which is operated by the Jet Propulsion Laboratory, California
Institute of Technology, under contract with the National Aeronautics and
Space Administration. This research
 also uses the Digitized Sky Survey (DSS) based
on photographic data obtained using Oschin Schmidt Telescope on
Palomar Mountain and The UK Schmidt Telescope and produced at the
Space Telescope Science Institute.

We thank M. Balcells for the stimulating discussion about the 
merger scenario for TBB galaxies,  V. Reshetnikov for his comments on
UGC\,10205, and F. Combes for her suggestions in her referee report.

\end{acknowledgements}


\begin{thebibliography}{}

\bibitem[2001]{agu} Aguerri, J. A. L., Balcells, M., \& Peletier, R. F. 2001, 
A\&A, 373, 786

\bibitem[1995]{and} Andredakis, Y. C., Peletier, R., \& Balcells M. 1995, MNRAS, 275, 874

\bibitem[2002]{am02} Athanassoula, E., \& Misiriotis, A. 2002, MNRAS, 330, 35

\bibitem[2003]{bal} Balcells, M. 2003, priv. communication

\bibitem[1994]{bp} Balcells, M., \& Peletier, R. 1994, AJ, 107, 135

\bibitem[1992]{bar} Barnes, J. E. 1992, ApJ, 331, 699

\bibitem[1992]{bar99} Barnes, J. E. 1999, Galaxy Interactions,
in Galaxy Dynamics, ASP Conf. Ser. 182,
ed.\ D. R. Merritt, M. Valluri, \& J. A. Sellwood, 463

\bibitem[1994]{bd} Barteldrees, A., \& Dettmar, R.-J. 1994, A\&AS, 103, 475

\bibitem[1989]{bel} Bell, M., \& Whitmore, B. C. 1989, ApJS, 70, 139

\bibitem[2000]{bb00} Bendo, G. J., \& Barnes, J. E. 2000, MNRAS, 316, 315

\bibitem[2003]{bahf03} Berentzen, I., Athanassoula, E., Heller, C. H., \&
Fricke, K. J. 2003, MNRAS, 341, 343

\bibitem[1985]{bin} Binney, J., \& Petrou, M. 1985, MNRAS, 214, 449

\bibitem[2002]{bc02} Bournaud, F., \& Combes, F. 2002 A\&A, 392, 83

\bibitem[1998]{bur} Bureau, M. 1998, Bars in Edge-On Spiral Galaxies,
Ph.D. Thesis, The Australian National University

\bibitem[1999]{bf} Bureau, M., \& Freeman, K. C. 1999, AJ, 118, 126

\bibitem[2000]{com99} Combes, F. 2000,
Bulge Formation, 
in Building Galaxies: from the Primordial Universe to the Present,
Proceedings of Rencontres de Moriond, 
ed.\ F. Hammer, T. X. Thuan,
V. Cayatte, B. Guiderdoni, \& J. Tran Thanh Van,
413

\bibitem[2003]{com03} Combes, F. 2003, priv. communication

\bibitem[1990]{com90} Combes, F., Debbasch, F., Friedli, D., \& Pfenniger, D. 1990,
A\&A, 233, 82

\bibitem[2001]{con2001} Conselice, C. 2001,
http://heritage.stsci.edu/2001/23/supple\-mental.html

\bibitem[1996]{cou} Courteau, S., de Jong, R. S., \& Broils, A. H. 1996, ApJ, 457, L73

\bibitem[1989]{det89} Dettmar, R.-J. 1989,
Box- and Peanut-Shaped Bulges of Disk Galaxies,
in The World of Galaxies, 
ed.\ H. G. Corwin, \& L. Bottinelli (Springer), 229

\bibitem[1990a]{db90a} Dettmar, R.-J., \& Barteldrees, A. 1990a,
Substructure and asymmetries in bulges,
in ESO/CTIO Workshop on Bul\-ges of Galaxies,
ed.\ B. J. Jarvis, \& D. M. Terndrup (ESO), 255

\bibitem[1999]{dl} Dettmar, R.-J., \& L\"utticke, R. 1999,
Do some bulges result from merging\,?,
in ASP Conf. Ser. 165, The Third Stromlo Symposium: The Galactic Halo, 
ed.\ B. K. Gibson, T. S. Axelrod,  \& M. E. Putman, 95

\bibitem[1948]{vau} de Vaucouleurs, G. 1948, Ann. d'Astrophys., 11, 247

\bibitem[1991]{rc3} de Vaucouleurs, G., de Vaucouleurs, A., Corwin, Jr., H. G.,
Buta, R. J., \& Fouqu\'e, P. 1991,
Third Reference Catalogue of Bright Galaxies (RC3) (Springer)

\bibitem[1996]{el96} Elmegreen, B. G., Elmegreen, D. M., Chromey, F. R.,
Hasselbacher, D. A., \& Bissell, B. A. 1996, AJ, 111, 2233

\bibitem[1985]{ee85} Elmegreen, B. G., \& Elmegreen, D. M. 1985, ApJ, 288, 438

\bibitem[1995]{ee95} Elmegreen, D. M., \& Elmegreen, B. G. 1995, ApJ, 445, 591

\bibitem[2003]{erw03} Erwin, P. 2003, priv. communication

\bibitem[2003]{es03} Erwin, P., \& Sparke L. S. 2003, ApJS, 146, 299

\bibitem[1999]{EPI} ESO Press Information 1999, 
http://www.eso.org\-/outreach\-/press-rel\-/pr-1999\-/phot-20-99.html

\bibitem[1994]{fis} Fisher, D., Illingworth, G., \& Franx, M. 1994,
AJ, 107, 160

\bibitem[2001]{fnsq} Font, A. S., Navarro, J. F., Stadel, J., Quinn T. 2001,
ApJ 563, 1

\bibitem[1993]{fb93} Friedli, D., \& Benz, W. 1993, A\&A, 268, 65

\bibitem[1993]{gar} Garcia, A. M. 1993, A\&AS, 100, 47

\bibitem[2001]{gb} Garcia-Burillo, S. 2001, priv. communication

\bibitem[1990]{ger} Gerin, M., Combes, F., \& Athanassoula, E. 1990, A\&A, 230, 37

\bibitem[1993]{hpn93} Hasan, H., Pfenniger, D., \& Norman, C. 1993, ApJ, 409, 91

\bibitem[1993]{her} Hernquist, L. 1993, ApJ, 409, 548

\bibitem[1989]{hq} Hernquist, L., \& Quinn, P. J. 1989, ApJ, 342, 1

\bibitem[1997]{hc97} Huang, S., \& Carlberg, R. G. 1997, ApJ, 480, 503

\bibitem[1993]{km} Kemp, S. N., \& Meaburn, J. 1993, A\&A, 274, 19

\bibitem[1995]{kui} Kuijken, K., \& Merrifield, M. R. 1995, ApJ, 433, L13

\bibitem[1996]{lue96} L\"utticke, R. 1996, Fl\"a\-chen\-pho\-to\-met\-ri\-sche
Un\-ter\-su\-chun\-gen zu
Eigen\-schaf\-ten von ``box/peanut''-bul\-ges in ``edge-on'' Schei\-ben\-ga\-la\-xi\-en,
Diploma Thesis, Ruhr-Uni\-ver\-si\-t\"at Bo\-chum

\bibitem[1999]{luediss} L\"utticke, R. 1999, Box and Peanut Shaped Bulges,
Ph.D. Thesis, Ruhr-Uni\-ver\-si\-t\"at Bo\-chum

\bibitem[1999]{ld99} L\"utticke, R., \& Dettmar, R.-J. 1999,
A New Class of Bulges,
in The Formation of Galactic Bulges, 
ed. C. M. Carollo, H. C. Ferguson, \& R. F. G. Wyse 
(Cambridge University Press), 119

\bibitem[2000a]{lue2000a} L\"utticke, R., Dettmar, R.-J., \& Pohlen, M. 2000a,
A\&AS, 145, 405 (Paper I)

\bibitem[2000b]{lue2000b} L\"utticke, R., Dettmar, R.-J., \& Pohlen, M. 2000b,
A\&A, 362, 435 (Paper II)

\bibitem[1985]{may} May, A., van Albada, T. S., \& Norman, C. A. 1985, MNRAS, 214, 131

\bibitem[1999]{mar99} Marlin, D., \&  Hadley, B. 1999,
Observational Evidence of Interactions in Bright, Nearby Galaxies,
in ASP Conf. Ser. 182,
ed.\ D. R. Merritt, M. Valluri, \& J. A. Sellwood, 445

\bibitem[1999]{mer} Merrifield, M., \& Kuijken, K. 1999, A\&A, 345, L47

\bibitem[1995]{mih} Mihos, J. C., Walker, I. R., Hernquist, L., de Oliveira, C. M., \&
Bolte, M. 1995, ApJ, 447, L87

\bibitem[1987]{nog} Noguchi, M. 1987, MNRAS, 228, 635

\bibitem[1996]{nsh96} Norman, C. A., Sellwood, J. A., \& Hasan, H. 1996, ApJ, 462, 114

\bibitem[1990]{oth90} Otha, K., Hamabe, M., \& Wakamatsu, K.-I. 1990, ApJ, 357, 71

\bibitem[2002a]{pags} Patsis, P. A., Athanassoula, E., Grosbol, P., \& Skokos, C. 2002a, MNRAS, 335, 1049

\bibitem[2002b]{psa02} Patsis, P. A., Skokos, C., \& Athanassoula, E. 2002b,
MNRAS, 337, 578

\bibitem[1996]{pb} Peletier, R., \& Balcells, M. 1996, AJ, 111, 2238

\bibitem[1999]{pfe99} Pfenniger, D. 1999, Ap\&SS, 269, 149

\bibitem[2000]{pfe2000} Pfenniger, D. 2000, 
Evolution Time-scales in the Hubble Sequence,
in ASP Conf. Ser. 197,
ed.\ F. Combes, G. A. Mamon, \& V. Charmandaris,  413

\bibitem[2001]{pohdiss} Pohlen, M. 2001,  Structures of Galactic
Stellar Disks, Ph.D. Thesis,
Ruhr-Uni\-ver\-si\-t\"at Bo\-chum

\bibitem[2000]{poh} Pohlen, M., Dettmar, R.-J., L\"utticke, R., \& Schwarzkopf, U.,
2000, A\&AS, 144, 405

\bibitem[1991]{rah} Raha, N., Sellwood, J. A., James, R. A., \& Kahn, F. D. 1991,
Nature, 352, 411

\bibitem[2003]{res03} Reshetnikov, V. P. 2003, priv. communication

\bibitem[1999]{re99} Reshetnikov, V. P., \& Evstigneeva, E. A. 1999,
Astronomy Reports, Volume 43, Issue 6, 367

\bibitem[1988]{row} Rowley, G. 1988, ApJ, 331, 124

\bibitem[1992]{sch} Schweizer, F., \& Seizer, P. 1992, AJ, 104, 1039

\bibitem[1999]{sm99} Sellwood, J. A., \& Moore, E. M. 1999, ApJ, 510, 125

\bibitem[1993]{sha93b} Shaw, M. A. 1993, A\&A, 280, 33

\bibitem[1995]{sha95} Shaw, M. A., Axon, D. J., Probst, R., \& Gatley, I. 1995, MNRAS, 274, 369

\bibitem[1990]{sh90} Shaw, M. A., Dettmar, R.-J., \& Barteldrees, A. 1990, A\&A, 240, 36

\bibitem[2003]{ss03} Shen, J., \& Sellwood, J. A. 2003, ApJ, in press (astro-ph/0310194)

\bibitem[1999]{ski} Skiff, B. 1999, priv. communication

\bibitem[1996]{vB} van den Bergh, S., Abraham, R., Ellis, R. S., 
Tanvir, N. R., 
Santiago, B. X., \& Glazebrook, K. G. 1996, AJ, 112, 359

\bibitem[2000]{vd2000} van Driel, W., Arnaboldi, M., Combes, F., \& Sparke, L. S.
2000, A\&AS, 141, 385

\bibitem[1997]{veg} Vega, J. S., Corsini, E. M., Pizzella, A., \& Bertola, F. 1997, A\&A,
 324, 485

\bibitem[1999]{vw99} Velazquez, H., \& White, S. D. M. 1999, MNRAS, 304, 254

\bibitem[2001a]{vdk} Vergani, D., Dettmar, R.-J., \& Klein, U. 2001a,
The HI kinematics of a Thick Boxy Bulge Galaxy: NGC\,1055,
in Dwarf Galaxies and their Environment,
ed.\ K.S. de Boer, R.-J. Dettmar, \& U. Klein (Shaker Verlag), 255

\bibitem[2001c]{vdkag} Vergani, D., Dettmar, R.-J., \& Klein, U. 2001c,
Multi-Wavelength Studies of Merging Bulge Galaxies,
in Astronomische Gesellschaft Abstract Series, Vol. 18., 553

\bibitem[2001b]{vpld} Vergani, D., Pohlen, M., L\"utticke, R., \& Dettmar, R.-J.
2001b, Thick Boxy Bulges in NIR, in Dwarf Galaxies and their Environment,
ed.\ K.S. de Boer, R.-J. Dettmar, \& U. Klein (Shaker Verlag), 251 

\bibitem[1977]{vv} Vorontsov-Velyaminov, B. A. 1977, A\&AS, 28, 1

\bibitem[1996]{wmh} Walker, I. R., Mihos, J. C., \& Hernquist, L. 1996, ApJ, 460, 121

\bibitem[1988]{whi} Whitmore, B. C., \& Bell, M. 1988, ApJ, 324, 741

\bibitem[1990]{wh90} Whitmore, B. C., Lucas, R. A., McElroy, D. B., Steiman-Cameron, T. Y., Sackett, P. D., \& Olling, R. P. 1990, AJ, 100, 1489

\bibitem[1999]{zg99} Zaritsky, D., \& Gonzalez, A. H. 1999, PASP, 111, 1508

\bibitem[1997]{zr97} Zaritsky, D., \& Rix, H.-W. 1997, ApJ, 477, 118

\end{thebibliography}
\end{document}